\documentclass[aps,reprint,showpacs,superscriptaddress,
footinbib,citeautoscript]{revtex4-1}

\usepackage{graphicx}

\usepackage[utf8]{inputenc}
\usepackage{csquotes}
\usepackage[american]{babel}
\usepackage[T1]{fontenc}
\usepackage{enumerate}
\usepackage{mdwlist}
\usepackage[activate=normal]{pdfcprot}
\usepackage{bbding}
\usepackage{color}
\usepackage{amsmath}
\usepackage{eqnarray,amsmath}
\usepackage{amssymb}
\usepackage{amsmath}
\usepackage{amsfonts}
\usepackage{mathrsfs}
\usepackage[titletoc,title]{appendix}
\usepackage{bm}
\usepackage{dcolumn}
\usepackage{color}
\usepackage[colorlinks=true,citecolor=blue]{hyperref}
\hypersetup{colorlinks=true,citecolor=blue,linkcolor=red
,urlcolor=blue}

\newcommand{\beq}{\begin{equation}}
\newcommand{\eeq}{\end{equation}}
\newcommand{\beqa}{\begin{eqnarray}}
\newcommand{\eeqa}{\end{eqnarray}}

\begin{document}

\title{Chiral Hall effect in strained Weyl semimetals}
\author{Shiva Heidari}
\affiliation{School of Physics, Institute for Research in Fundamental Sciences, IPM, Tehran, 19395-5531, Iran}
\author{ Reza Asgari}
\email{asgari@ipm.ir}
\affiliation{School of Physics, Institute for Research in Fundamental Sciences, IPM, Tehran, 19395-5531, Iran}
\affiliation{School of Nano Science, Institute for Research in Fundamental Sciences, IPM, Tehran, 19395-5531, Iran}
\affiliation{ARC Centre of Excellence in Future Low-Energy Electronics Technologies,  UNSW Node,  Sydney 2052,  Australia}
\begin{abstract}
In this paper, the chiral Hall effect of strained Weyl semimetals without any external magnetic field is proposed. Electron-phonon coupling emerges in the low-energy fermionic sector through a pseudogauge potential. We show that, by using chiral kinetic theory, the chiral Hall effect appears as a response to a real time-varying electric field in the presence of structural distortion and
it causes spatial chirality and charges separation in a Weyl system. We also show that the coupling of the electrons to acoustic phonons as a gapless excitation leads to emerging an optical absorption peak at $\omega=\omega_{el}$, where $\omega_{el}$ is defined as a characteristic frequency associated with the pseudomagnetic field. We also propose the strain-induced planar Hall effect as another transport signature of the chiral-anomaly equation.
\end{abstract}
\maketitle

\section{Introduction}\label{sec:intro}
Nontrivial topological structure of Weyl semimetals (WSMs) has opened new avenues for discovery of unprecedented anomaly induced transport phenomena \citep{RevModPhys.90.015001,doi:10.1146/annurev-conmatphys-033117-054129,doi:10.1146/annurev-conmatphys-031016-025225,doi:10.1146/annurev-conmatphys-031016-025458,doi:10.1146/annurev-conmatphys-031113-133841,PhysRevB.84.235126,RevModPhys.88.035005}. The spin nondegenerate band topology of WSMs could be described by accidental band crossing at some certain points in the Brillouin zone when either time-reversal or spatial inversion symmetry has been broken. The topological notion of WSMs can be understood as the topological protection of Berry curvature flux which according to Gauss's theorem is quantized to integer values known as Chern numbers or monopole charges. Such monopole charges can be only added or removed in pairs \citep{NIELSEN1983389}. Weyl fermions will be effectively stabilized by a nonzero axial four-vector $b_\mu=(b_0,\bm{b})$, defined as the separation of Weyl nodes in momentum ($2 b$) and energy space ($2 b_0$) even in the absence of any additional point-group symmetries.

Notably, the electromagnetic response of WSMs is modified by the topological structure of the system. It leads to emergence of an axionic term in the electromagnetic response as a signature of the so-called chiral anomaly $\partial_\mu J^\mu_5=\frac{e^2}{2 \pi^2} \bm{E} \cdot \bm{B}$ \citep{PhysRevB.86.115133,PhysRevLett.111.027201,bertlmann2000anomalies}. The most prominent consequence of the chiral anomaly is a positive longitudinal magnetoconductance confirmed by several pieces of experimental evidence in Dirac and WSMs \citep{He_2014,Xiong_2015,Hirschberger_2016,Li_2016,Zhang_2016,Liang_2014}. Another unique transport signature of chiral anomaly is the planar Hall effect which has been theoretically predicted and experimentally observed in WSMs \citep{Nandy_2017,Kumar_2018,Li_2018}. 

Lattice distortion, on the other hand, couples to the low-energy electronic sector through elastic gauge fields with opposite signs at different nodal points \citep{PhysRevLett.115.177202,ilan2019pseudo,PhysRevX.6.041021}.
 Lattice distortion can lead to spatial and/or time variation of Weyl nodes positions ($b_\mu \rightarrow b_\mu (r,t)$) which in turn induces pseudomagnetic and pseudoelectric fields defined as $\bm{B}_5=\nabla \times \bm{b}$ and $\bm{E}_5=(\partial_t \bm{b}- v_F \nabla b_0)$, respectively. The spatial dependence of $\bm{b}$ can be achieved by applying strain \citep{PhysRevLett.115.177202,PhysRevB.92.165131}, inhomogeneous magnetization \citep{PhysRevB.87.235306,PhysRevB.90.134409}, and at the interface of two WSMs with different $\bm{b}$ vectors  \citep{PhysRevX.6.041046,PhysRevX.7.041026,PhysRevB.95.214103}. Moreover, boundaries always host a finite axial magnetic field $\bm{B}_5$ where $\bm{b}$ jumps from zero to a finite value or vice versa \citep{PhysRevX.6.041046, ilan2019pseudo}.

Coupling of electrons to elastic gauge fields not only induces novel chiral anomaly in WSMs $\partial_\mu J^\mu_5=\frac{e^2}{6 \pi^2} (3 \bm{E} \cdot \bm{B} + \bm{E}_5 \cdot \bm{B}_5)$, but also results in nonconservation of total charge $\partial_\mu J^\mu=\frac{e^2}{2 \pi^2} (\bm{E} \cdot \bm{B}_5 + \bm{E}_5 \cdot \bm{B})$  \citep{PhysRevB.87.235306}. The novel chiral anomaly and nonconservation of charge possess great potential for uncovering new physical signatures of electron-phonon coupling in strained WSMs.  The basic paradigm of these new phenomena is the chiral pseudomagnetic effect \citep{PhysRevX.6.041021,Sumiyoshi_2016,Huang_2017}.

  Electron-phonon coupling in the form of elastic gauge fields also contributes to the modification of effective acoustic \citep{PhysRevB.92.165131,PhysRevLett.115.177202,Cortijo_2016,Chernodub_2019} and optical \citep{PhysRevB.100.165427} phonon dynamics, nonequilibrium chiral magnetic effects \citep{Cortijo_PRB}, and emergent excitations in the fermionic spectrum \citep{van_der_Wurff_2019}. Moreover, the electron-phonon coupling may lead to the planar Hall effect in strained WSMs even in the absence of a real magnetic field \citep{ghosh2019chirality}. Furthermore, a strain field could derive an electric current and that the effect was dictated by a second-class Chern invariant~\cite{Zhou_2013}. Strained WSMs have also received much attention due to the experimental realizations  \citep{Jia_2019,Peri_2019,Kamboj_2019}.
 
In the present paper, we propose a \textit{chiral Hall effect} as another consequence of electron-phonon coupling in strained WSMs. We demonstrate that a current is induced as a response to a real time-varying electric field $\bm{E}$ in the presence of $\bm{B}_5$. Since the induced current is perpendicular to both $\bm{E}$ and $\bm{B}_5$, we name it the chiral Hall effect (CHE). The CHE may lead to spatial chirality separation in the case of zero axial chemical potential $\mu_5(=\mu_+-\mu_-$), and both chirality and charge separation in the case of nonzero $\mu_5$. Charge and chirality separation has been studied in the presence of a real magnetic field \citep{PhysRevB.98.165205, PhysRevB.98.205149}. The parabolic Hall effect as a special type of chiral separation has also been proposed \citep{breitkreiz2019parabolic}. Here, it is shown that the chiral Hall effect in strained WSMs leads to spatial chirality and charge separation without any external magnetic field. The emergence of the strain-induced transverse chiral Hall current is our main finding. This work thus demonstrates the chiral Hall effect in the context of WSMs.

We also show that an extra intraband absorption peak in the optical conductivity of doped WSMs appears as a result of the pseudomagnetic field. Furthermore, the presence of the anomaly term ($\bm{E} \cdot \bm{B}_5$) may be caused by a strain-induced enhancement of the longitudinal electric conductivity. It brings about an in-plane transport signature through anomaly-induced charge pumping from the boundaries to the bulk. The \textit{planar Hall effect} owing to the lattice distortion as an exquisite transport phenomenon emerges when both electric and pseudomagnetic fields are located on the same plane. This is another significant result of this paper.

The rest of this paper is organized as follows. We commence with a description of our theoretical formalism in Sec. \ref{sec2}, followed by details of the mechanism of chiral kinetic theory in WSMs and the nonequilibrium distribution function. Analytical expressions for the longitudinal and transverse total current and chiral current in the presence of a deformed lattice are provided in this section. 
In Sec. \ref{sec3} we summarize our main findings.

\section{Model and theoretical formalism} \label{sec2}
We consider a time-reversal symmetry broken WSM under lattice deformation. A low-energy effective Hamiltonian in a continuum limit near nodal points is given by
\begin{equation}
{\cal H}=\hbar v(\bm{q}+\chi \bm{A}^{el})\cdot \bm{\sigma}
\end{equation}
where $v$ is the Fermi velocity and $\chi(=\pm 1)$ denotes the chirality associated with nodal points are located at $\bm{b}$ and $-\bm{b}$. The pseudomagnetic field is obtained from $\bm{B}^{el}=\bm{\nabla} \times \bm{A}^{el}$ where the axial gauge potential $\bm{A}^{el}$ is the spatially dependent part of $\bm{b}$. For our purpose, it is enough to consider a kind of lattice distortion which leads to a nonzero pseudomagnetic field. We shall investigate the response of strained WSMs to a real time-varying electric field in the framework of semiclassical chiral kinetic theory. 

Chiral kinetic theory is a topologically modified semiclassical Boltzmann formalism to describe the behavior of Weyl fermions for a finite chemical potential.
We are interested in doped WSMs where the Fermi level $\mu$ crosses the conduction band. In the regime of $\hbar \omega \ll \mu \text{ and } k_B T \ll \mu $, the absorption process is governed by intraband transitions with a negligible contribution of interband transitions. Chiral kinetic theory obeys semiclassical description where the wave packet of Bloch electrons has a Gaussian shape and spreads in real and momentum spaces. Most importantly, WSMs reveal nontrivial and anomalous transport properties through semiclassical Boltzmann formalism where two chiral Fermi surfaces enclose a nonzero flux of Berry curvature in momentum space.

\subsection{Semiclassical Boltzmann transport theory in WSMs}

 We would like to investigate the response of strained Weyl system under a weak oscillating electric field with frequency $\omega$ and wavevector $\bm{q}$
\beq
\bm{E}(\bm{r},t)=(E_x,E_y,E_z) e^{i \bm{q}\cdot \bm{r} -i \omega t}
\eeq 
where $E_i$ could be a real or complex number depending on the linearly or circularly polarized light, respectively. In the semiclassical framework, the equations of motion are given by
\beq \label{EOM}
\begin{split}
& {\cal D} \dot{\bm{x}}^\chi=\bm{v}^\chi -e \bm{\Omega}^\chi \times  \bm{E}-\chi (\bm{v}^\chi \cdot \bm{\Omega}^\chi)  \bm{B}^{el}\\
& {\cal D} \dot{\bm{k}}^\chi= e \bm{E}+\chi \bm{v}^\chi \times \bm{B}^{el}+e \chi ( \bm{E} \cdot  \bm{B}^{el}) \bm{\Omega}^\chi
\end{split}
\eeq
where we use $\bm{\Omega}^\chi=\chi \bm{\Omega}=\chi \frac{\hat{k}}{2 |k|^2}$ for our WSM model. The factor ${\cal D}=1-\chi \bm{\Omega}^\chi \cdot \bm{B}^{el}=1-\frac{\hat{k}\cdot \bm{B}^{el}}{2 |k|^2}$ accounts for the modification of phase space satisfies Liouville's theorem \citep{Duval_2006}. Note that pseudofields couple to chirality as a charge-less degree of freedom unlike real electromagnetic fields which couple to electron charge in the low-energy Hamiltonian.

The energy of Weyl fermions in the presence of a weak effective pseudomagnetic field $\bm{B}^{el}$ acquires a term owing to the intrinsic magnetic moment, therefore, the dispersion relation is modified as (we set $\hbar=1$)
\beq
\epsilon_k=v k (1+\bm{B}^{el} \cdot \bm{\Omega})
\eeq
and the corresponding velocity is given by 
\beq \label{velocity}
\bm{v}=\bm{\nabla}_k \epsilon_k=v \hat{k} [1-2 \bm{B}^{el} \cdot \bm{\Omega}]+v \bm{B}^{el} (\hat{k} \cdot \bm{\Omega}).
\eeq 
We note that the chirality drops out of the group velocity, { i.e.} both modes are right-moving in the bulk. This is in contrast with the condition of applying a real magnetic field where each mode disperses oppositely \citep{ilan2019pseudo, PhysRevX.6.041021}. 

The nonequilibrium distribution function $f_\chi (t,p,x)$ for a given chirality $\chi$ is obtained as a solution of the semiclassical Boltzmann equation  
\begin{equation} \label{boltz}
\dfrac{\partial f_\chi}{\partial t}+\dot{\mathbf{x}} \cdot \dfrac{\partial f_\chi}{\partial \mathbf{x}}+\dot{\mathbf{k}} \cdot \dfrac{\partial f_\chi}{\partial \mathbf{k}}=I_{coll}(f_\chi)
\end{equation}
where $I_{coll}(f_\chi)$ is the collision integral. We focus on the low temperatures $k_B T \ll \mu $ and finite frequencies where the \textit{collisionless limit} $\omega \tau \gg 1$ ($\tau$ is the shortest relaxation time) is valid \citep{Roy_2016}. Therefore the RHS of Eq. \ref{boltz} vanishes.
For the sake of simplicity, we consider the linear-response regime where corrections to equilibrium distribution function $f^{(eq)}$ are linear in $\bm{E}$.  We therefore choose $f_\chi=f^{(eq)}+\frac{\partial f_0}{\partial \epsilon} \delta f_\chi e^{-i \omega +i \bm{q} \cdot \bm{r}}$ and obtain
\begin{equation}\label{new}
\begin{split}
&(1-\frac{\hat{k} \cdot  \bm{B}^{el}}{2 |k|^2}) \partial_t \delta f_\chi+[\bm{v} -\chi(\bm{v} \cdot \bm{\Omega}_\chi) \bm{B}^{el}] \cdot \bm{\partial}_r \delta f_\chi+ \\ &  + [e \bm{E}-\chi \bm{v} \times \bm{B}^{el}-\chi e (\bm{E} \cdot \bm{B}^{el}) \bm{\Omega}^\chi] \cdot \bm{\partial}_k f^{(eq)} - \\ &  -\chi (\bm{v} \times \bm{B}^{el}) \cdot \bm{\partial}_k \delta f_\chi=0
\end{split}
\end{equation}
Substituting $\bm{v}$ from  Eq. \ref{velocity} into Eq. \ref{new} and considering linear terms in $\bm{E}$, the above expression becomes 
\begin{equation}
\begin{split}
i [(1+\kappa)& \omega -v (\bm{q} \cdot \hat{k})] \delta f^\chi +\chi v (\hat{k} \times  \bm{B}^{el}) \cdot \bm{\partial}_k \delta f^\chi \\ & = e v [\bm{E} \cdot \hat{k} -(\bm{E} \cdot \bm{B}^{el})(\bm{\Omega} \cdot \hat{k})] (1+2\kappa)
\end{split}
\end{equation}
where we define $\kappa= \frac{\hat{k} \cdot \bm{B}^{el}}{2 |k|^2}$. Making use of the spherical coordinate and considering pseudomagnetic field along the $z$ axis we obtain

\begin{equation}
\begin{split}
\chi \omega^{el} & \partial_\varphi \delta f_\chi +i [(1+\kappa)\omega-v (\hat{k} \cdot \bm{q})] \delta f_\chi \\ &= e v [\hat{k} \cdot \bm{E} - (\bm{E} \cdot \bm{B}^{el})(\bm{\Omega} \cdot \hat{k})]
\end{split}
\end{equation}
where $\omega^{el}=\nu {B}^{el}/k$ is the cyclotron frequency of massless fermions in the pseudomagnetic field. In a long-wavelength limit ($q \ll 1$), the analysis simplifies significantly because one can neglect the wavevector $q$ dependence on $\delta f_\chi$. We make use of the standard parametrization of vectors in spherical coordinates to decompose $\delta f_\chi$ into harmonics as $\delta f_\chi=\delta f_0+\delta f_\chi^+ e^{i \varphi}+\delta f_\chi^- e^{-i \varphi}$ and obtain the following solutions which are linear in $\bm{E}$
\begin{equation} \label{boltzsol}
\begin{split}
& \delta f_\chi^0=-i e v \dfrac{ (\cos \theta- \alpha)}{\omega (1+\alpha \cos \theta)} \ E \cdot \hat{z}, \\
& \delta f_\chi^+=-i e v \dfrac{\sin \theta}{(1+\alpha \cos \theta) \omega+\chi  \omega^{el}} E_-,\\
& \delta f_\chi^-=-i e v \dfrac{ \sin \theta}{(1+\alpha \cos \theta) \omega-\chi  \omega^{el}} E_+.
\end{split}
\end{equation}
Here $E_+=\frac{E_x+i E_y}{2}$, $E_-=\frac{E_x-i E_y}{2}$, $\kappa=\alpha \cos \theta$ with $\alpha=\frac{B^{el}}{2 k^2}$ and accordingly $\omega^{el}=2 \alpha |\mu|$.

\subsection{Optical current density}
\subsubsection{The electric field is perpendicular to the pseudomagnetic field: $ \mathbf{E} \perp \mathbf{B}^{el} $ }
In this stage, once the nonequilibrium distribution function is obtained, we can calculate total and chiral current densities through $J_\mu=\sum_\chi  J^\chi_\mu $ and  $J^5_\mu=\sum_\chi \chi J^\chi_\mu$, respectively. The current associated with each node is given by
\begin{equation} \label{chiralcur}
\bm{J}^\chi=-e\int \dfrac{d^3 k}{(2 \pi)^3}  f_\chi(t,k,x) {\cal D} \dot{\bm{x}}_\chi 
\end{equation}
where
\begin{equation} \label{DX}
{\cal D} \dot{\bm{x}}_i^\chi=v \hat{k}_i (1-2 \alpha \cos \theta)+v \alpha^2 \cos \theta \ \hat{n} + e \bm{E} \times \bm{\Omega}^\chi .
\end{equation}
Here $\hat{n}$ is the unit vector pointing along the pseudomagnetic field $\bm{B}^{el}$. \ $\bm{J}^\chi$ can be further simplified by using Eq. \ref{DX} and keeping terms up to linear order on $\bm{E}$, the current of each node reads 
\begin{equation}
J^\chi_{i,\perp}=-e v  \int \dfrac{d^3 k}{(2 \pi)^3} \hat{k}_i (1-2 \kappa) \delta f_\chi \dfrac{\partial f^{(eq)}}{\partial \epsilon_k} .
\end{equation}
At sufficiently low temperatures $\partial f^{eq}/ \partial \epsilon_k=-\delta(\mu-\epsilon_k)$, then the above integral simplifies to
\begin{equation}
J^\chi_{i,\perp}= \dfrac{e \mu^2}{(2 \pi)^3 v^2} \int_{-1}^{1} \int_{0}^{2 \pi} du \ d\varphi \ \hat{k}_i (1-2 \alpha u) \delta f_\chi
\end{equation}
with the change of variable $u=\cos \theta$. Considering the pseudomagnetic field as a small perturbation, we can evaluate the integral in the regime of small $\alpha$ where pseudo-Landau-level quantization is unimportant. In general, a Weyl semimetal consists of an even number of Weyl points \citep{NIELSEN1983389}. We study the case of two Weyl points, therefore, our results can be easily generalized to other Weyl systems with more than two nodes. 

Once the total current is obtained, optical conductivity tensor which is defined by 
\begin{equation}
J_i=\sigma_{ij} E_j
\end{equation}
can be calculated. The longitudinal optical conductivity is therefore given by
\begin{equation} \label{sigma1}
\hat{\sigma}_{xx}=\hat{\sigma}_{yy}=-\dfrac{ie \mu^2 }{3 v \pi^2}  \omega (\dfrac{1}{\omega^2-\omega_{el}^2}+\dfrac{3}{5} \alpha^2 \omega^2 \dfrac{\omega^2+\omega_{el}^2}{(\omega^2-\omega_{el}^2)^3})
\end{equation}
and the transverse optical response per node $\chi$ is
\begin{equation}
\hat{\sigma}_{xy}^\chi=- \hat{\sigma}_{yx}^\chi=\dfrac{\chi e^2 \mu_\chi^2 }{6 \pi^2 v} \omega_{el} (\dfrac{1}{\omega^2-\omega_{el}^2}+\dfrac{7}{5} \alpha^2 \omega^2 \dfrac{\omega^2+\frac{3}{7} \omega_{el}^2}{(\omega^2-\omega_{el}^2)^3})
\end{equation}
where $\mu_{\chi}$ is the chemical potential at node $\chi$. The hat label denotes that these optical conductivity expressions are complex-valued numbers. To specify the real and imaginary part of $\hat{\sigma}$, we add a small quantity $i \gamma$ to the frequency ($\omega \rightarrow \omega-i \gamma $). The $\gamma$ parameter prevents the divergence of the conductivity. After summing over chirality, the off-diagonal elements of the optical conductivity tensor cancel each other in the $\mu_+=\mu_-$ case. In the following, we demonstrate that a spectacular feature occurs for $\mu_+ \neq \mu_-$. It should be pointed out that our results are valid in ac mode, i.e. in the frequency interval $0<\omega \ll 2 \mu $, which induces intraband transitions apart from the dc Drude response. Using Eq. \ref{sigma1}, the general form of optical response to the time-varying electric field would be
\begin{figure}
\includegraphics[width=6.5 cm,height=12 cm]{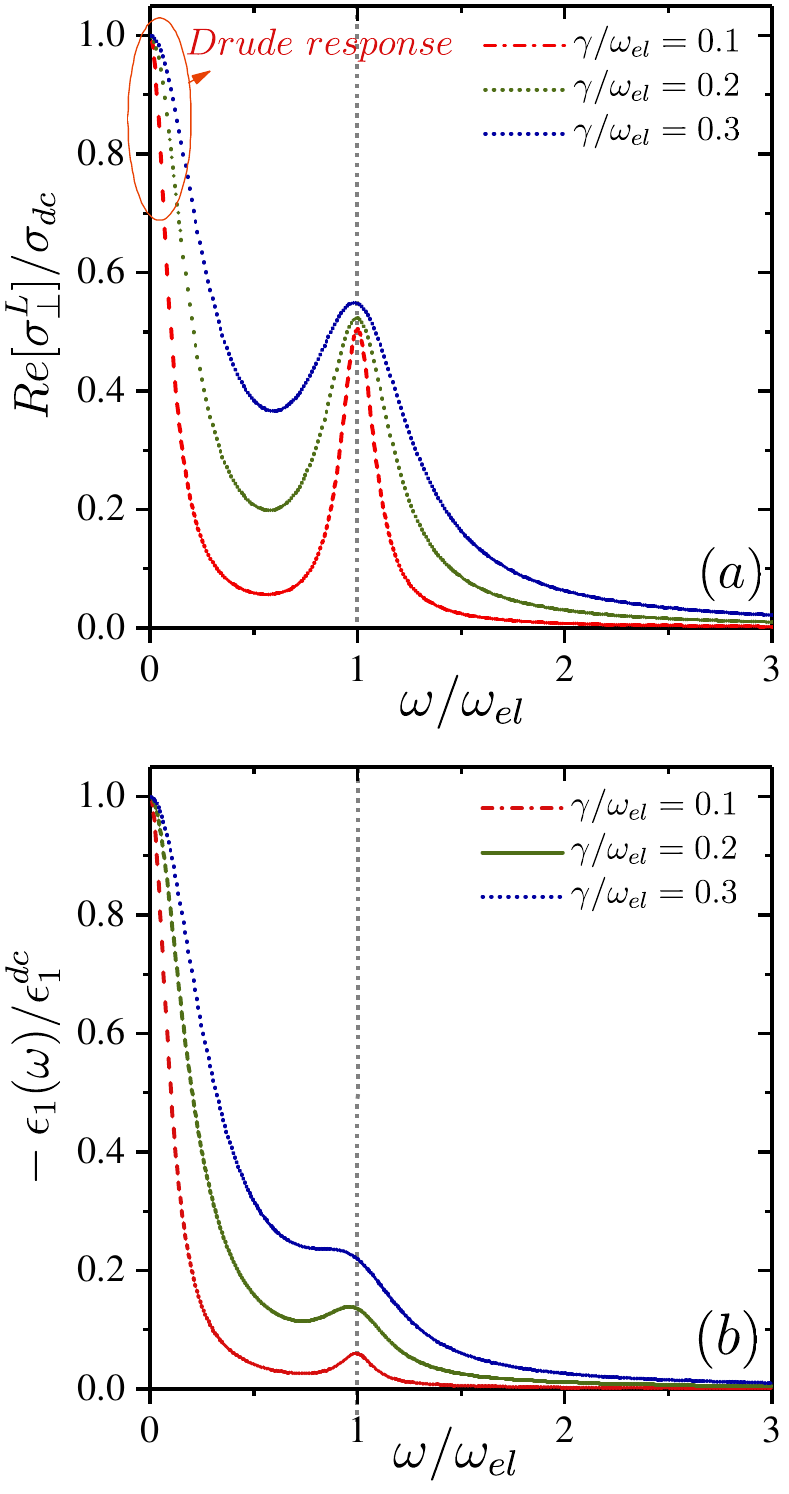}
\caption{(Color online)(a): Real part of intraband optical conductivity as a function of frequency.  The zero-frequency response corresponds to the conventional Drude peak decreases monotonically by frequency. The presence of the pseudomagnetic field leads to the emergence of an absorption peak at $\omega=\omega_{el}$ ($\omega,\omega_{el} \ll \mu $). We consider different $\gamma$ values in units of $\omega_{el}$ and show that the width of the peak is broadened with a rising $\gamma$ parameter. (b): Real part of dielectric function versus frequency shows a moderate peak of around $\omega=\omega_{el}$ that stems from the pseudomagnetic field. The slope of the dielectric curve around $\omega=\omega_{el}$ becomes smooth by $\gamma$ increment. We set $\alpha=0.01$ and $\omega_{el}=2 \alpha |\mu|$} \label{fig1}
\end{figure}

\begin{equation}
Re[\sigma_{\perp}^L(\omega)]=\sigma_{dc} \delta(\omega) +Re[\hat{\sigma}_{xx}]
\end{equation}
where $\sigma_{dc}=e \mu^2/3 \pi^2 v \gamma$ determines the maximum of the Drude response at zero frequency and $\gamma$ is a small broadening parameter. The second part, originates from the pseudomagnetic field and is valid for the range of $0<\omega \ll 2 \mu $.
Figure \ref{fig1}(a) illustrates the low-frequency optical response in the presence of strain. The longitudinal optical conductivity shows two absorption peaks centered at $\omega=0$ and $\omega=\omega_{el}$ associated with the conventional Drude response and the strain-induced absorption peak, respectively. This novel peak leads us to infer that the optical conductivity $\sigma_{\perp}$ is highly enhanced by chiral electrons when the external frequency $\omega$ approaches $\omega_{el}$. It is noteworthy that this chirality accumulation can be detected by the imbalanced absorbance of the circularly polarized light~\citep{Hosur_2015}.

The dielectric function is an important physical quantity to explore. In general, it is a complex-valued function where its imaginary part determines the amount of absorption inside a medium. Most importantly, including the quantum interlayer contributions leads to increasing the imaginary part of the dielectric function. It turns out that the effects of quantum mechanics are vital for systems in which interlayer transitions play an important role. There is a well-known connection between the real part of the dielectric function, $\epsilon_1$, and the imaginary part of optical conductivity 
\begin{equation}
\epsilon_1(\omega)=\epsilon_b-\frac{4\pi Im [\sigma_{\perp}^L(\omega)]}{\omega}.
\end{equation}
Here $\epsilon_b$ is the high-frequency dielectric constant of the WSM. It should be recalled that the real part of a dielectric function or the imaginary part of a conductivity describes the reflection of light.

Figure \ref{fig1}(b) represents the real part of the dielectric function versus normalized frequency. It is useful to note that at the onset of transparency (plasmon frequency) we have $\epsilon_1(\omega)$=0. Since this function is always negative, which leads the metal to exhibit reflectivity, there is no plasmon mode in the range of parameters considered.

In contrast to the total charge current, the chiral (axial) current will not vanish even if $\mu_+=\mu_-$
\begin{equation}
\begin{split}
J^5_i=& \sum_{\chi} \chi \sigma_{ij}^\chi E_j=\\  =[\sum_\chi & \dfrac{e^2 \mu_\chi^2 }{6 \pi^2 v} \omega_{el} (\dfrac{1}{\omega^2-\omega_{el}^2}+\dfrac{7}{5} \alpha^2 \omega^2 \dfrac{\omega^2+\frac{3}{7} \omega_{el}^2}{(\omega^2-\omega_{el}^2)^3})] \epsilon^{ij} E_j.
\end{split}
\end{equation}
\begin{figure}
\includegraphics[width=8.5 cm,height=9 cm]{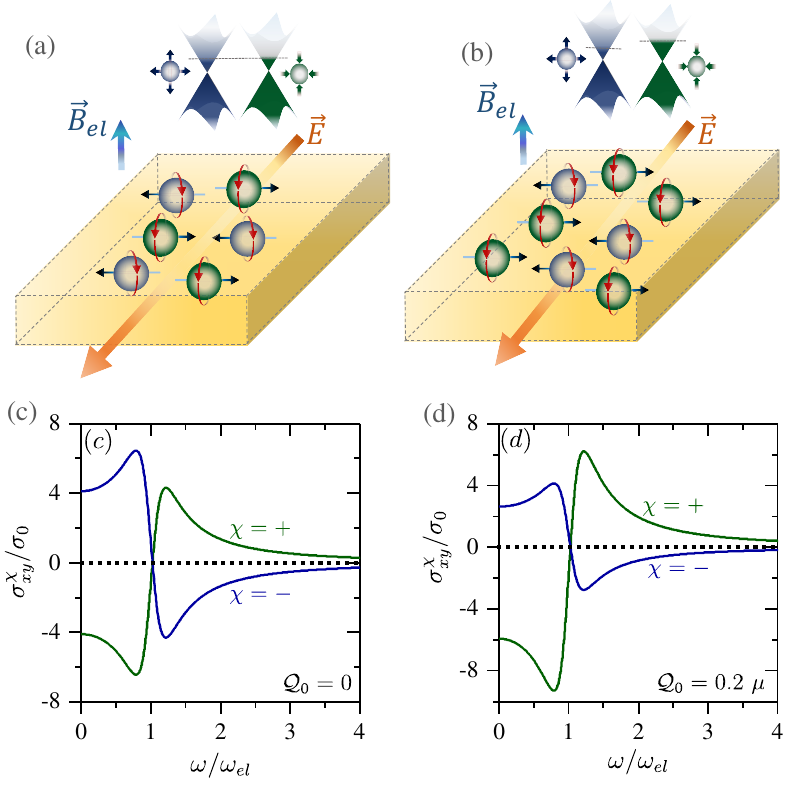}
\caption{(Color online) (a), (b): Schematic representation of CHE in the bulk. (a,c): In the case of $\mu_+=\mu_-$, the number of right- and left-handed fermions is the same but they disperse in opposite direction. It leads to a zero net charge current. Accordingly, only chirality separation occurs in this situation. (b,d): A moderate inversion-symmetry breaking causes $2{\cal Q}_0=\mu_+ - \mu_- \neq 0$. Therefore, in contrast with panels (a,c) both charge and chirality separation arise in a strained Weyl system. We set $\gamma / \omega_{el}=0.2$, $\alpha=0.01$, $\mu / \omega_{el}=100$ and $\sigma_0=4 e^2 \pi^2 \omega_{el}/v$.} \label{fig2}
\end{figure}
\begin{figure}
\includegraphics[width=9 cm]{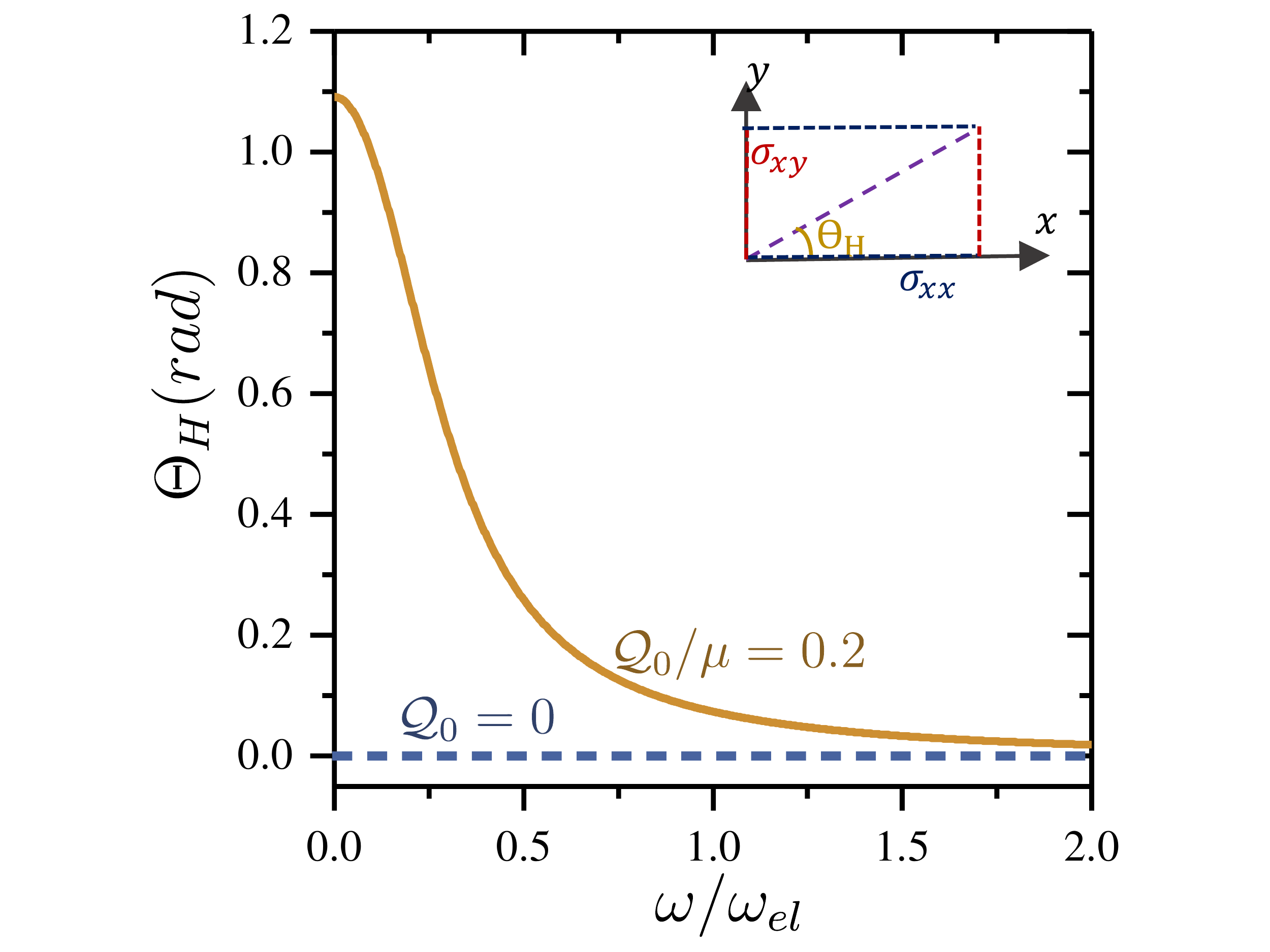}
\caption{(Color online): Bulk Hall angle as a function of frequency. We set $\gamma / \omega_{el}=0.2$, $\alpha=0.01$.} \label{fig3}
\end{figure}
This strain-induced axial current perpendicular to both the ordinary time-varying electric field and the pseudomagnetic field is interpreted as the chiral Hall effect \citep{PhysRevD.91.025011}. Note that this current vanishes in the absence of a pseudomagnetic field. In the case of $\mu_+=\mu_-$, the current of right-handed fermions ($\chi=+$) is the same in magnitude but in the opposite direction as the left-handed fermions ($\chi=-$) [Fig. \ref{fig2}(a)]. It is very important to note that $\bm{J^{5}}$ is a current of chirality without flowing a net charge. Carriers with opposite chirality diffuse in opposite directions leading to chirality separation in space [Fig. \ref{fig2}(c)]. This phenomenon is analogous to the spin Hall effect where spin-orbit interaction causes spin separation.

The prerequisite condition to get nonzero Hall conductivity is fulfilled by $\mu_+ \neq \mu_-$. Such an inversion symmetry breaking can be provided by e.g. momentum-independent spin-orbit interaction \cite{Zyuzin_2012}. Generically, in the case of $\mu_+ \neq \mu_- $, the chemical potential at the plus node ($\chi=+$) is pushed up by an amount of ${\cal Q}_0$ while $\mu$ at the minus node is pushed down by the same amount. In this situation, the currents of right-handed and left-handed fermions are again in the opposite direction but this time with different magnitudes [Fig. \ref{fig2}(b) and Fig. \ref{fig2}(d)]. Accordingly, the currents corresponding to each chirality do not compensate each other and the system develops a net charge. As a result, CHE leads to both charge and chirality separation at the same time. 
Based on the results of Fig. \ref{fig2}, chiral Hall current changes sign at $\omega/\omega_{el}=1$. Therefore the chirality of accumulated charges at each side of the sample will be reversed at $\omega=\omega_{el}$.

In addition, a physical quantity that is related to the Nernst parameter is the bulk Hall angle. Having calculated conductivity tensor, the bulk Hall angle can also be obtained as
\begin{equation}
 \Theta_H=\tan^{-1} \left( Re\left[\dfrac{\hat{\sigma}_{xy}(\omega)}{\hat{\sigma}_{xx}(\omega)}\right] \right).
\end{equation}
In the dc limit, $\tan(\Theta_H)\!=\!-2 e {\cal Q}_0\omega_{el}/ (\gamma \mu)$ showing that the transverse Hall conductivity $\sigma_{xy}$ vanishes if either $\omega_{el}$ or ${\cal Q}_0$ being zero. When the character of chiral particle changes from $\mu<0$ to $\mu>0$, Hall angle also changes from positive to negative. Fig. \ref{fig3} shows the Hall angle as a function of frequency. In either case of $\mu_+=\mu_-$ or ${\cal Q}_0=0$, the only nonzero optical response is along the electric field (longitudinal $\sigma_{xx}$) and Hall angle vanishes throughout all the frequency range. On the other hand, transverse optical response ($\sigma_{xy}$) becomes non-zero for ${\cal Q}_0\neq 0$ which leads to a finite Hall angle. The Hall angle decreases monotonically by frequency increment meaning that the longitudinal response dominates rather than the transverse one.

\subsubsection{The electric field is parallel to the pseudomagnetic field: $ \mathbf{E} \parallel \mathbf{B}^{el} $}
 
In a similar manner of the ordinary chiral magnetic effect (CME) in the presence of a real $\bm{B}$ field, the pseudomagnetic field also generates a chiral pseudomagnetic effect. As a sharp contrast, the ordinary CME vanishes in equilibrium ($\mu_+ =\mu_-$) while chiral pseudomagnetic effect may exist \citep{PhysRevX.6.041046,PhysRevX.6.041021,ilan2019pseudo}. In the latter case, the left-moving modes near boundaries and the right-moving modes in bulk compensate each other after summing over the entire volume. There is an electrical charge moving between regions of opposite $\bm{B}^{el}$ via anomalous Hall effect. In particular, the sign of $\bm{B}^{el}$ changes near boundary regions where $b_\mu$ varies rapidly as a consequence of nodes annihilation. 

When an electric field $\bm{E}$ is applied parallel to the pseudomagnetic field, according to the novel anomaly equation, a charge-density imbalance between bulk and surface is induced \citep{PhysRevX.6.041021}
\begin{equation}
\delta \rho=\dfrac{e^2 \tau}{2 \pi^2} \bm{E} \cdot \bm{B}^{el}.
\end{equation}

Here, $\tau$ is the electron relaxation time. This charge-density imbalance can relax through the processes which the right-moving modes in bulk scatter back to the left-moving modes near the boundaries.
The induced bulk electron density gets disturbed among all the empty bulk states above Fermi energy and leads to a small shift in chemical potential as $\mu \rightarrow \mu +\delta \mu$. In the limit of  $ k_B T \ll \delta \mu \ll \mu $, the shift in chemical potential can be approximated as
\begin{equation}
\delta \mu \approx \dfrac{2 \pi^2 v^3}{\mu^2} \delta \rho,
\end{equation}
which is valid in the semiclassical limit where pseudo-Landau level quantization is unimportant. Accordingly, the population of the two right-moving modes in bulk rises with the same amount $\mu_\pm=\mu_\pm^0+\delta \mu$.

In the collisionless limit, we suppose the particle path is straight and not affected by a circular motion owing to the pseudomagnetic field. Therefore relaxation time $\tau$ is field independent. The corresponding conductivity associated with charge pumping from boundary to bulk is called the chiral torsional effect (CTE)
\begin{equation} \label{CTE}
\sigma_{CTE}=\dfrac{e^4 v^3}{8 \pi^3} \dfrac{\tau}{\mu^2} (B^{el})^2
\end{equation}
which leads to $J_{CTE}^z=\sigma_{CTE} E_z $ that cannot be captured by chiral kinetic theory.
The main consequence of the above equation is the enhancement of bulk transport current originated directly from the novel chiral anomaly.

The parallel current $J^\chi_\parallel $ in the chiral kinetic theory framework is given by
\begin{equation}
\begin{split}
& J_\parallel^\chi=ev \int \dfrac{d^3 k}{(2 \pi)^3} (-\dfrac{\partial f^{(eq)}}{\partial \epsilon_k})  (\cos \theta (1-2 \alpha \cos \theta)\\ &+ \alpha^2 \cos^2 \theta) \delta f_\chi^0 -e \int \dfrac{d^3 k}{(2 \pi)^3} f^{(eq)} v \hat{k}_i (1-2 \alpha \cos \theta).
\end{split}
\end{equation}
From this expression, we can read off the parallel component of conductivity tensor
\begin{equation}
\sigma_{zz}=-\dfrac{ie^2 \mu^2}{4 \pi^2 v \omega} \sum_{\chi} \int_{-1}^{1} du \dfrac{(u (1-2 \alpha u)+\alpha^2 u^2)(u-\alpha)}{(1+\alpha u)}.
\end{equation}
As before, we expand integrand up to the second-order in $\alpha$. After integrating over $u$ and collecting the terms, the longitudinal optical conductivity along the pseudomagnetic field is given by
\begin{equation}
\sigma_{zz}=\dfrac{\sigma_{dc}}{1+i\omega \tau} (1+\dfrac{24}{5} \alpha^2).
\end{equation}\\
The correction term is proportional to $\alpha^2 \propto (B^{el})^2$ having a positive contribution to the longitudinal conductivity in a strained WSM system. Similar discussions of strain-induced enhancement of electric conductivity in WSMs have also been carried out by other groups \citep{Landsteiner_2018,PhysRevX.6.041021,PhysRevX.6.041046}.
Finally, we remark that the net longitudinal conductivity is given by $\tilde{\sigma}_{zz}=\sigma_{zz}+\sigma_{CTE}$ where $\sigma_{zz}$ comes from chiral kinetic calculus and $\sigma_{CTE}$ stems from the novel chiral anomaly. 

\section{Conclusion} \label{sec3}
In summary, we have considered strained and doped Weyl semimetals in the presence of an external electric field. We have explored the chiral Hall effect, focusing on intraband transitions as a consequence of coupling light to the free carriers. Chiral kinetic theory is utilized to calculate the elements of optical conductivity tensor. Structural deformation together with the time-varying electric field leads to the charge and chirality separation in WSMs when a moderate inversion symmetry has been broken. In addition, the pseudomagnetic field induces a longitudinal conductivity peak at the cyclotron frequency of massless fermions. Furthermore, by employing chiral kinetic theory we have introduced a generalized formula for anomaly-induced electric conductivity in the strained WSMs.

Imbalanced absorption of the left- and right-handed circularly polarized light is a technique to determine the surface chiral charge of a pristine system. Our findings here can be explored by making use of this experimental technique. Our proposal on the charge and chirality separation in strained WSMs provides a broad outlook for future studies and potential applications of emerging Weyl semimetals.

\section{Acknowledgement}
This work is supported by Iran Science Elites Federation.  

\bibliography{CHE1}

\begin{thebibliography}{50}%
\makeatletter
\providecommand \@ifxundefined [1]{%
 \@ifx{#1\undefined}
}%
\providecommand \@ifnum [1]{%
 \ifnum #1\expandafter \@firstoftwo
 \else \expandafter \@secondoftwo
 \fi
}%
\providecommand \@ifx [1]{%
 \ifx #1\expandafter \@firstoftwo
 \else \expandafter \@secondoftwo
 \fi
}%
\providecommand \natexlab [1]{#1}%
\providecommand \enquote  [1]{``#1''}%
\providecommand \bibnamefont  [1]{#1}%
\providecommand \bibfnamefont [1]{#1}%
\providecommand \citenamefont [1]{#1}%
\providecommand \href@noop [0]{\@secondoftwo}%
\providecommand \href [0]{\begingroup \@sanitize@url \@href}%
\providecommand \@href[1]{\@@startlink{#1}\@@href}%
\providecommand \@@href[1]{\endgroup#1\@@endlink}%
\providecommand \@sanitize@url [0]{\catcode `\\12\catcode `\$12\catcode
  `\&12\catcode `\#12\catcode `\^12\catcode `\_12\catcode `\%12\relax}%
\providecommand \@@startlink[1]{}%
\providecommand \@@endlink[0]{}%
\providecommand \url  [0]{\begingroup\@sanitize@url \@url }%
\providecommand \@url [1]{\endgroup\@href {#1}{\urlprefix }}%
\providecommand \urlprefix  [0]{URL }%
\providecommand \Eprint [0]{\href }%
\providecommand \doibase [0]{http://dx.doi.org/}%
\providecommand \selectlanguage [0]{\@gobble}%
\providecommand \bibinfo  [0]{\@secondoftwo}%
\providecommand \bibfield  [0]{\@secondoftwo}%
\providecommand \translation [1]{[#1]}%
\providecommand \BibitemOpen [0]{}%
\providecommand \bibitemStop [0]{}%
\providecommand \bibitemNoStop [0]{.\EOS\space}%
\providecommand \EOS [0]{\spacefactor3000\relax}%
\providecommand \BibitemShut  [1]{\csname bibitem#1\endcsname}%
\let\auto@bib@innerbib\@empty
\bibitem [{\citenamefont {Armitage}\ \emph {et~al.}(2018)\citenamefont
  {Armitage}, \citenamefont {Mele},\ and\ \citenamefont
  {Vishwanath}}]{RevModPhys.90.015001}%
  \BibitemOpen
  \bibfield  {author} {\bibinfo {author} {\bibfnamefont {N.~P.}\ \bibnamefont
  {Armitage}}, \bibinfo {author} {\bibfnamefont {E.~J.}\ \bibnamefont {Mele}},
  \ and\ \bibinfo {author} {\bibfnamefont {A.}~\bibnamefont {Vishwanath}},\
  }\href {\doibase 10.1103/RevModPhys.90.015001} {\bibfield  {journal}
  {\bibinfo  {journal} {Rev. Mod. Phys.}\ }\textbf {\bibinfo {volume} {90}},\
  \bibinfo {pages} {015001} (\bibinfo {year} {2018})}\BibitemShut {NoStop}%
\bibitem [{\citenamefont
  {Burkov}(2018)}]{doi:10.1146/annurev-conmatphys-033117-054129}%
  \BibitemOpen
  \bibfield  {author} {\bibinfo {author} {\bibfnamefont {A.}~\bibnamefont
  {Burkov}},\ }\href {\doibase 10.1146/annurev-conmatphys-033117-054129}
  {\bibfield  {journal} {\bibinfo  {journal} {Annual Review of Condensed Matter
  Physics}\ }\textbf {\bibinfo {volume} {9}},\ \bibinfo {pages} {359} (\bibinfo
  {year} {2018})},\ \Eprint
  {http://arxiv.org/abs/https://doi.org/10.1146/annurev-conmatphys-033117-054129}
  {https://doi.org/10.1146/annurev-conmatphys-033117-054129} \BibitemShut
  {NoStop}%
\bibitem [{\citenamefont {Hasan}\ \emph {et~al.}(2017)\citenamefont {Hasan},
  \citenamefont {Xu}, \citenamefont {Belopolski},\ and\ \citenamefont
  {Huang}}]{doi:10.1146/annurev-conmatphys-031016-025225}%
  \BibitemOpen
  \bibfield  {author} {\bibinfo {author} {\bibfnamefont {M.~Z.}\ \bibnamefont
  {Hasan}}, \bibinfo {author} {\bibfnamefont {S.-Y.}\ \bibnamefont {Xu}},
  \bibinfo {author} {\bibfnamefont {I.}~\bibnamefont {Belopolski}}, \ and\
  \bibinfo {author} {\bibfnamefont {S.-M.}\ \bibnamefont {Huang}},\ }\href
  {\doibase 10.1146/annurev-conmatphys-031016-025225} {\bibfield  {journal}
  {\bibinfo  {journal} {Annual Review of Condensed Matter Physics}\ }\textbf
  {\bibinfo {volume} {8}},\ \bibinfo {pages} {289} (\bibinfo {year} {2017})},\
  \Eprint
  {http://arxiv.org/abs/https://doi.org/10.1146/annurev-conmatphys-031016-025225}
  {https://doi.org/10.1146/annurev-conmatphys-031016-025225} \BibitemShut
  {NoStop}%
\bibitem [{\citenamefont {Yan}\ and\ \citenamefont
  {Felser}(2017)}]{doi:10.1146/annurev-conmatphys-031016-025458}%
  \BibitemOpen
  \bibfield  {author} {\bibinfo {author} {\bibfnamefont {B.}~\bibnamefont
  {Yan}}\ and\ \bibinfo {author} {\bibfnamefont {C.}~\bibnamefont {Felser}},\
  }\href {\doibase 10.1146/annurev-conmatphys-031016-025458} {\bibfield
  {journal} {\bibinfo  {journal} {Annual Review of Condensed Matter Physics}\
  }\textbf {\bibinfo {volume} {8}},\ \bibinfo {pages} {337} (\bibinfo {year}
  {2017})},\ \Eprint
  {http://arxiv.org/abs/https://doi.org/10.1146/annurev-conmatphys-031016-025458}
  {https://doi.org/10.1146/annurev-conmatphys-031016-025458} \BibitemShut
  {NoStop}%
\bibitem [{\citenamefont {Vafek}\ and\ \citenamefont
  {Vishwanath}(2014)}]{doi:10.1146/annurev-conmatphys-031113-133841}%
  \BibitemOpen
  \bibfield  {author} {\bibinfo {author} {\bibfnamefont {O.}~\bibnamefont
  {Vafek}}\ and\ \bibinfo {author} {\bibfnamefont {A.}~\bibnamefont
  {Vishwanath}},\ }\href {\doibase 10.1146/annurev-conmatphys-031113-133841}
  {\bibfield  {journal} {\bibinfo  {journal} {Annual Review of Condensed Matter
  Physics}\ }\textbf {\bibinfo {volume} {5}},\ \bibinfo {pages} {83} (\bibinfo
  {year} {2014})},\ \Eprint
  {http://arxiv.org/abs/https://doi.org/10.1146/annurev-conmatphys-031113-133841}
  {https://doi.org/10.1146/annurev-conmatphys-031113-133841} \BibitemShut
  {NoStop}%
\bibitem [{\citenamefont {Burkov}\ \emph {et~al.}(2011)\citenamefont {Burkov},
  \citenamefont {Hook},\ and\ \citenamefont {Balents}}]{PhysRevB.84.235126}%
  \BibitemOpen
  \bibfield  {author} {\bibinfo {author} {\bibfnamefont {A.~A.}\ \bibnamefont
  {Burkov}}, \bibinfo {author} {\bibfnamefont {M.~D.}\ \bibnamefont {Hook}}, \
  and\ \bibinfo {author} {\bibfnamefont {L.}~\bibnamefont {Balents}},\ }\href
  {\doibase 10.1103/PhysRevB.84.235126} {\bibfield  {journal} {\bibinfo
  {journal} {Phys. Rev. B}\ }\textbf {\bibinfo {volume} {84}},\ \bibinfo
  {pages} {235126} (\bibinfo {year} {2011})}\BibitemShut {NoStop}%
\bibitem [{\citenamefont {Chiu}\ \emph {et~al.}(2016)\citenamefont {Chiu},
  \citenamefont {Teo}, \citenamefont {Schnyder},\ and\ \citenamefont
  {Ryu}}]{RevModPhys.88.035005}%
  \BibitemOpen
  \bibfield  {author} {\bibinfo {author} {\bibfnamefont {C.-K.}\ \bibnamefont
  {Chiu}}, \bibinfo {author} {\bibfnamefont {J.~C.~Y.}\ \bibnamefont {Teo}},
  \bibinfo {author} {\bibfnamefont {A.~P.}\ \bibnamefont {Schnyder}}, \ and\
  \bibinfo {author} {\bibfnamefont {S.}~\bibnamefont {Ryu}},\ }\href {\doibase
  10.1103/RevModPhys.88.035005} {\bibfield  {journal} {\bibinfo  {journal}
  {Rev. Mod. Phys.}\ }\textbf {\bibinfo {volume} {88}},\ \bibinfo {pages}
  {035005} (\bibinfo {year} {2016})}\BibitemShut {NoStop}%
\bibitem [{\citenamefont {Nielsen}\ and\ \citenamefont
  {Ninomiya}(1983)}]{NIELSEN1983389}%
  \BibitemOpen
  \bibfield  {author} {\bibinfo {author} {\bibfnamefont {H.}~\bibnamefont
  {Nielsen}}\ and\ \bibinfo {author} {\bibfnamefont {M.}~\bibnamefont
  {Ninomiya}},\ }\href {\doibase https://doi.org/10.1016/0370-2693(83)91529-0}
  {\bibfield  {journal} {\bibinfo  {journal} {Physics Letters B}\ }\textbf
  {\bibinfo {volume} {130}},\ \bibinfo {pages} {389 } (\bibinfo {year}
  {1983})}\BibitemShut {NoStop}%
\bibitem [{\citenamefont {Zyuzin}\ and\ \citenamefont
  {Burkov}(2012)}]{PhysRevB.86.115133}%
  \BibitemOpen
  \bibfield  {author} {\bibinfo {author} {\bibfnamefont {A.~A.}\ \bibnamefont
  {Zyuzin}}\ and\ \bibinfo {author} {\bibfnamefont {A.~A.}\ \bibnamefont
  {Burkov}},\ }\href {\doibase 10.1103/PhysRevB.86.115133} {\bibfield
  {journal} {\bibinfo  {journal} {Phys. Rev. B}\ }\textbf {\bibinfo {volume}
  {86}},\ \bibinfo {pages} {115133} (\bibinfo {year} {2012})}\BibitemShut
  {NoStop}%
\bibitem [{\citenamefont {Vazifeh}\ and\ \citenamefont
  {Franz}(2013)}]{PhysRevLett.111.027201}%
  \BibitemOpen
  \bibfield  {author} {\bibinfo {author} {\bibfnamefont {M.~M.}\ \bibnamefont
  {Vazifeh}}\ and\ \bibinfo {author} {\bibfnamefont {M.}~\bibnamefont
  {Franz}},\ }\href {\doibase 10.1103/PhysRevLett.111.027201} {\bibfield
  {journal} {\bibinfo  {journal} {Phys. Rev. Lett.}\ }\textbf {\bibinfo
  {volume} {111}},\ \bibinfo {pages} {027201} (\bibinfo {year}
  {2013})}\BibitemShut {NoStop}%
\bibitem [{\citenamefont {Bertlmann}(2000)}]{bertlmann2000anomalies}%
  \BibitemOpen
  \bibfield  {author} {\bibinfo {author} {\bibfnamefont {R.~A.}\ \bibnamefont
  {Bertlmann}},\ }\href@noop {} {\emph {\bibinfo {title} {Anomalies in quantum
  field theory}}},\ Vol.~\bibinfo {volume} {91}\ (\bibinfo  {publisher} {Oxford
  University Press},\ \bibinfo {year} {2000})\BibitemShut {NoStop}%
\bibitem [{\citenamefont {He}\ \emph {et~al.}(2014)\citenamefont {He},
  \citenamefont {Hong}, \citenamefont {Dong}, \citenamefont {Pan},
  \citenamefont {Zhang}, \citenamefont {Zhang},\ and\ \citenamefont
  {Li}}]{He_2014}%
  \BibitemOpen
  \bibfield  {author} {\bibinfo {author} {\bibfnamefont {L.~P.}\ \bibnamefont
  {He}}, \bibinfo {author} {\bibfnamefont {X.~C.}\ \bibnamefont {Hong}},
  \bibinfo {author} {\bibfnamefont {J.~K.}\ \bibnamefont {Dong}}, \bibinfo
  {author} {\bibfnamefont {J.}~\bibnamefont {Pan}}, \bibinfo {author}
  {\bibfnamefont {Z.}~\bibnamefont {Zhang}}, \bibinfo {author} {\bibfnamefont
  {J.}~\bibnamefont {Zhang}}, \ and\ \bibinfo {author} {\bibfnamefont {S.~Y.}\
  \bibnamefont {Li}},\ }\href {\doibase 10.1103/PhysRevLett.113.246402}
  {\bibfield  {journal} {\bibinfo  {journal} {Phys. Rev. Lett.}\ }\textbf
  {\bibinfo {volume} {113}},\ \bibinfo {pages} {246402} (\bibinfo {year}
  {2014})}\BibitemShut {NoStop}%
\bibitem [{\citenamefont {Xiong}\ \emph {et~al.}(2015)\citenamefont {Xiong},
  \citenamefont {Kushwaha}, \citenamefont {Liang}, \citenamefont {Krizan},
  \citenamefont {Hirschberger}, \citenamefont {Wang}, \citenamefont {Cava},\
  and\ \citenamefont {Ong}}]{Xiong_2015}%
  \BibitemOpen
  \bibfield  {author} {\bibinfo {author} {\bibfnamefont {J.}~\bibnamefont
  {Xiong}}, \bibinfo {author} {\bibfnamefont {S.~K.}\ \bibnamefont {Kushwaha}},
  \bibinfo {author} {\bibfnamefont {T.}~\bibnamefont {Liang}}, \bibinfo
  {author} {\bibfnamefont {J.~W.}\ \bibnamefont {Krizan}}, \bibinfo {author}
  {\bibfnamefont {M.}~\bibnamefont {Hirschberger}}, \bibinfo {author}
  {\bibfnamefont {W.}~\bibnamefont {Wang}}, \bibinfo {author} {\bibfnamefont
  {R.~J.}\ \bibnamefont {Cava}}, \ and\ \bibinfo {author} {\bibfnamefont
  {N.~P.}\ \bibnamefont {Ong}},\ }\href {\doibase 10.1126/science.aac6089}
  {\bibfield  {journal} {\bibinfo  {journal} {Science}\ }\textbf {\bibinfo
  {volume} {350}},\ \bibinfo {pages} {413} (\bibinfo {year}
  {2015})}\BibitemShut {NoStop}%
\bibitem [{\citenamefont {Hirschberger}\ \emph {et~al.}(2016)\citenamefont
  {Hirschberger}, \citenamefont {Kushwaha}, \citenamefont {Wang}, \citenamefont
  {Gibson}, \citenamefont {Liang}, \citenamefont {Belvin}, \citenamefont
  {Bernevig}, \citenamefont {Cava},\ and\ \citenamefont
  {Ong}}]{Hirschberger_2016}%
  \BibitemOpen
  \bibfield  {author} {\bibinfo {author} {\bibfnamefont {M.}~\bibnamefont
  {Hirschberger}}, \bibinfo {author} {\bibfnamefont {S.}~\bibnamefont
  {Kushwaha}}, \bibinfo {author} {\bibfnamefont {Z.}~\bibnamefont {Wang}},
  \bibinfo {author} {\bibfnamefont {Q.}~\bibnamefont {Gibson}}, \bibinfo
  {author} {\bibfnamefont {S.}~\bibnamefont {Liang}}, \bibinfo {author}
  {\bibfnamefont {C.~A.}\ \bibnamefont {Belvin}}, \bibinfo {author}
  {\bibfnamefont {B.~A.}\ \bibnamefont {Bernevig}}, \bibinfo {author}
  {\bibfnamefont {R.~J.}\ \bibnamefont {Cava}}, \ and\ \bibinfo {author}
  {\bibfnamefont {N.~P.}\ \bibnamefont {Ong}},\ }\href {\doibase
  10.1038/nmat4684} {\bibfield  {journal} {\bibinfo  {journal} {Nature
  Materials}\ }\textbf {\bibinfo {volume} {15}},\ \bibinfo {pages} {1161}
  (\bibinfo {year} {2016})}\BibitemShut {NoStop}%
\bibitem [{\citenamefont {Li}\ \emph {et~al.}(2016)\citenamefont {Li},
  \citenamefont {Kharzeev}, \citenamefont {Zhang}, \citenamefont {Huang},
  \citenamefont {Pletikosi{\'{c}}}, \citenamefont {Fedorov}, \citenamefont
  {Zhong}, \citenamefont {Schneeloch}, \citenamefont {Gu},\ and\ \citenamefont
  {Valla}}]{Li_2016}%
  \BibitemOpen
  \bibfield  {author} {\bibinfo {author} {\bibfnamefont {Q.}~\bibnamefont
  {Li}}, \bibinfo {author} {\bibfnamefont {D.~E.}\ \bibnamefont {Kharzeev}},
  \bibinfo {author} {\bibfnamefont {C.}~\bibnamefont {Zhang}}, \bibinfo
  {author} {\bibfnamefont {Y.}~\bibnamefont {Huang}}, \bibinfo {author}
  {\bibfnamefont {I.}~\bibnamefont {Pletikosi{\'{c}}}}, \bibinfo {author}
  {\bibfnamefont {A.~V.}\ \bibnamefont {Fedorov}}, \bibinfo {author}
  {\bibfnamefont {R.~D.}\ \bibnamefont {Zhong}}, \bibinfo {author}
  {\bibfnamefont {J.~A.}\ \bibnamefont {Schneeloch}}, \bibinfo {author}
  {\bibfnamefont {G.~D.}\ \bibnamefont {Gu}}, \ and\ \bibinfo {author}
  {\bibfnamefont {T.}~\bibnamefont {Valla}},\ }\href {\doibase
  10.1038/nphys3648} {\bibfield  {journal} {\bibinfo  {journal} {Nature
  Physics}\ }\textbf {\bibinfo {volume} {12}},\ \bibinfo {pages} {550}
  (\bibinfo {year} {2016})}\BibitemShut {NoStop}%
\bibitem [{\citenamefont {Zhang}\ \emph {et~al.}(2016)\citenamefont {Zhang},
  \citenamefont {Xu}, \citenamefont {Belopolski}, \citenamefont {Yuan},
  \citenamefont {Lin}, \citenamefont {Tong}, \citenamefont {Bian},
  \citenamefont {Alidoust}, \citenamefont {Lee}, \citenamefont {Huang},
  \citenamefont {Chang}, \citenamefont {Chang}, \citenamefont {Hsu},
  \citenamefont {Jeng}, \citenamefont {Neupane}, \citenamefont {Sanchez},
  \citenamefont {Zheng}, \citenamefont {Wang}, \citenamefont {Lin},
  \citenamefont {Zhang}, \citenamefont {Lu}, \citenamefont {Shen},
  \citenamefont {Neupert}, \citenamefont {Hasan},\ and\ \citenamefont
  {Jia}}]{Zhang_2016}%
  \BibitemOpen
  \bibfield  {author} {\bibinfo {author} {\bibfnamefont {C.-L.}\ \bibnamefont
  {Zhang}}, \bibinfo {author} {\bibfnamefont {S.-Y.}\ \bibnamefont {Xu}},
  \bibinfo {author} {\bibfnamefont {I.}~\bibnamefont {Belopolski}}, \bibinfo
  {author} {\bibfnamefont {Z.}~\bibnamefont {Yuan}}, \bibinfo {author}
  {\bibfnamefont {Z.}~\bibnamefont {Lin}}, \bibinfo {author} {\bibfnamefont
  {B.}~\bibnamefont {Tong}}, \bibinfo {author} {\bibfnamefont {G.}~\bibnamefont
  {Bian}}, \bibinfo {author} {\bibfnamefont {N.}~\bibnamefont {Alidoust}},
  \bibinfo {author} {\bibfnamefont {C.-C.}\ \bibnamefont {Lee}}, \bibinfo
  {author} {\bibfnamefont {S.-M.}\ \bibnamefont {Huang}}, \bibinfo {author}
  {\bibfnamefont {T.-R.}\ \bibnamefont {Chang}}, \bibinfo {author}
  {\bibfnamefont {G.}~\bibnamefont {Chang}}, \bibinfo {author} {\bibfnamefont
  {C.-H.}\ \bibnamefont {Hsu}}, \bibinfo {author} {\bibfnamefont {H.-T.}\
  \bibnamefont {Jeng}}, \bibinfo {author} {\bibfnamefont {M.}~\bibnamefont
  {Neupane}}, \bibinfo {author} {\bibfnamefont {D.~S.}\ \bibnamefont
  {Sanchez}}, \bibinfo {author} {\bibfnamefont {H.}~\bibnamefont {Zheng}},
  \bibinfo {author} {\bibfnamefont {J.}~\bibnamefont {Wang}}, \bibinfo {author}
  {\bibfnamefont {H.}~\bibnamefont {Lin}}, \bibinfo {author} {\bibfnamefont
  {C.}~\bibnamefont {Zhang}}, \bibinfo {author} {\bibfnamefont {H.-Z.}\
  \bibnamefont {Lu}}, \bibinfo {author} {\bibfnamefont {S.-Q.}\ \bibnamefont
  {Shen}}, \bibinfo {author} {\bibfnamefont {T.}~\bibnamefont {Neupert}},
  \bibinfo {author} {\bibfnamefont {M.~Z.}\ \bibnamefont {Hasan}}, \ and\
  \bibinfo {author} {\bibfnamefont {S.}~\bibnamefont {Jia}},\ }\href {\doibase
  10.1038/ncomms10735} {\bibfield  {journal} {\bibinfo  {journal} {Nature
  Communications}\ }\textbf {\bibinfo {volume} {7}} (\bibinfo {year} {2016}),\
  10.1038/ncomms10735}\BibitemShut {NoStop}%
\bibitem [{\citenamefont {Liang}\ \emph {et~al.}(2014)\citenamefont {Liang},
  \citenamefont {Gibson}, \citenamefont {Ali}, \citenamefont {Liu},
  \citenamefont {Cava},\ and\ \citenamefont {Ong}}]{Liang_2014}%
  \BibitemOpen
  \bibfield  {author} {\bibinfo {author} {\bibfnamefont {T.}~\bibnamefont
  {Liang}}, \bibinfo {author} {\bibfnamefont {Q.}~\bibnamefont {Gibson}},
  \bibinfo {author} {\bibfnamefont {M.~N.}\ \bibnamefont {Ali}}, \bibinfo
  {author} {\bibfnamefont {M.}~\bibnamefont {Liu}}, \bibinfo {author}
  {\bibfnamefont {R.~J.}\ \bibnamefont {Cava}}, \ and\ \bibinfo {author}
  {\bibfnamefont {N.~P.}\ \bibnamefont {Ong}},\ }\href {\doibase
  10.1038/nmat4143} {\bibfield  {journal} {\bibinfo  {journal} {Nature
  Materials}\ }\textbf {\bibinfo {volume} {14}},\ \bibinfo {pages} {280}
  (\bibinfo {year} {2014})}\BibitemShut {NoStop}%
\bibitem [{\citenamefont {Nandy}\ \emph {et~al.}(2017)\citenamefont {Nandy},
  \citenamefont {Sharma}, \citenamefont {Taraphder},\ and\ \citenamefont
  {Tewari}}]{Nandy_2017}%
  \BibitemOpen
  \bibfield  {author} {\bibinfo {author} {\bibfnamefont {S.}~\bibnamefont
  {Nandy}}, \bibinfo {author} {\bibfnamefont {G.}~\bibnamefont {Sharma}},
  \bibinfo {author} {\bibfnamefont {A.}~\bibnamefont {Taraphder}}, \ and\
  \bibinfo {author} {\bibfnamefont {S.}~\bibnamefont {Tewari}},\ }\href
  {\doibase 10.1103/PhysRevLett.119.176804} {\bibfield  {journal} {\bibinfo
  {journal} {Phys. Rev. Lett.}\ }\textbf {\bibinfo {volume} {119}},\ \bibinfo
  {pages} {176804} (\bibinfo {year} {2017})}\BibitemShut {NoStop}%
\bibitem [{\citenamefont {Kumar}\ \emph {et~al.}(2018)\citenamefont {Kumar},
  \citenamefont {Guin}, \citenamefont {Felser},\ and\ \citenamefont
  {Shekhar}}]{Kumar_2018}%
  \BibitemOpen
  \bibfield  {author} {\bibinfo {author} {\bibfnamefont {N.}~\bibnamefont
  {Kumar}}, \bibinfo {author} {\bibfnamefont {S.~N.}\ \bibnamefont {Guin}},
  \bibinfo {author} {\bibfnamefont {C.}~\bibnamefont {Felser}}, \ and\ \bibinfo
  {author} {\bibfnamefont {C.}~\bibnamefont {Shekhar}},\ }\href {\doibase
  10.1103/PhysRevB.98.041103} {\bibfield  {journal} {\bibinfo  {journal} {Phys.
  Rev. B}\ }\textbf {\bibinfo {volume} {98}},\ \bibinfo {pages} {041103}
  (\bibinfo {year} {2018})}\BibitemShut {NoStop}%
\bibitem [{\citenamefont {Li}\ \emph {et~al.}(2018)\citenamefont {Li},
  \citenamefont {Wang}, \citenamefont {He}, \citenamefont {Wang},\ and\
  \citenamefont {Shen}}]{Li_2018}%
  \BibitemOpen
  \bibfield  {author} {\bibinfo {author} {\bibfnamefont {H.}~\bibnamefont
  {Li}}, \bibinfo {author} {\bibfnamefont {H.-W.}\ \bibnamefont {Wang}},
  \bibinfo {author} {\bibfnamefont {H.}~\bibnamefont {He}}, \bibinfo {author}
  {\bibfnamefont {J.}~\bibnamefont {Wang}}, \ and\ \bibinfo {author}
  {\bibfnamefont {S.-Q.}\ \bibnamefont {Shen}},\ }\href {\doibase
  10.1103/PhysRevB.97.201110} {\bibfield  {journal} {\bibinfo  {journal} {Phys.
  Rev. B}\ }\textbf {\bibinfo {volume} {97}},\ \bibinfo {pages} {201110}
  (\bibinfo {year} {2018})}\BibitemShut {NoStop}%
\bibitem [{\citenamefont {Cortijo}\ \emph {et~al.}(2015)\citenamefont
  {Cortijo}, \citenamefont {Ferreir\'os}, \citenamefont {Landsteiner},\ and\
  \citenamefont {Vozmediano}}]{PhysRevLett.115.177202}%
  \BibitemOpen
  \bibfield  {author} {\bibinfo {author} {\bibfnamefont {A.}~\bibnamefont
  {Cortijo}}, \bibinfo {author} {\bibfnamefont {Y.}~\bibnamefont
  {Ferreir\'os}}, \bibinfo {author} {\bibfnamefont {K.}~\bibnamefont
  {Landsteiner}}, \ and\ \bibinfo {author} {\bibfnamefont {M.~A.~H.}\
  \bibnamefont {Vozmediano}},\ }\href {\doibase 10.1103/PhysRevLett.115.177202}
  {\bibfield  {journal} {\bibinfo  {journal} {Phys. Rev. Lett.}\ }\textbf
  {\bibinfo {volume} {115}},\ \bibinfo {pages} {177202} (\bibinfo {year}
  {2015})}\BibitemShut {NoStop}%
\bibitem [{\citenamefont {Ilan}\ \emph {et~al.}(2019)\citenamefont {Ilan},
  \citenamefont {Grushin},\ and\ \citenamefont {Pikulin}}]{ilan2019pseudo}%
  \BibitemOpen
  \bibfield  {author} {\bibinfo {author} {\bibfnamefont {R.}~\bibnamefont
  {Ilan}}, \bibinfo {author} {\bibfnamefont {A.~G.}\ \bibnamefont {Grushin}}, \
  and\ \bibinfo {author} {\bibfnamefont {D.~I.}\ \bibnamefont {Pikulin}},\
  }\href {\doibase 10.1038/s42254-019-0121-8} {\bibfield  {journal} {\bibinfo
  {journal} {Nature Reviews Physics}\ ,\ \bibinfo {pages} {1}} (\bibinfo {year}
  {2019})}\BibitemShut {NoStop}%
\bibitem [{\citenamefont {Pikulin}\ \emph {et~al.}(2016)\citenamefont
  {Pikulin}, \citenamefont {Chen},\ and\ \citenamefont
  {Franz}}]{PhysRevX.6.041021}%
  \BibitemOpen
  \bibfield  {author} {\bibinfo {author} {\bibfnamefont {D.~I.}\ \bibnamefont
  {Pikulin}}, \bibinfo {author} {\bibfnamefont {A.}~\bibnamefont {Chen}}, \
  and\ \bibinfo {author} {\bibfnamefont {M.}~\bibnamefont {Franz}},\ }\href
  {\doibase 10.1103/PhysRevX.6.041021} {\bibfield  {journal} {\bibinfo
  {journal} {Phys. Rev. X}\ }\textbf {\bibinfo {volume} {6}},\ \bibinfo {pages}
  {041021} (\bibinfo {year} {2016})}\BibitemShut {NoStop}%
\bibitem [{\citenamefont {Shapourian}\ \emph {et~al.}(2015)\citenamefont
  {Shapourian}, \citenamefont {Hughes},\ and\ \citenamefont
  {Ryu}}]{PhysRevB.92.165131}%
  \BibitemOpen
  \bibfield  {author} {\bibinfo {author} {\bibfnamefont {H.}~\bibnamefont
  {Shapourian}}, \bibinfo {author} {\bibfnamefont {T.~L.}\ \bibnamefont
  {Hughes}}, \ and\ \bibinfo {author} {\bibfnamefont {S.}~\bibnamefont {Ryu}},\
  }\href {\doibase 10.1103/PhysRevB.92.165131} {\bibfield  {journal} {\bibinfo
  {journal} {Phys. Rev. B}\ }\textbf {\bibinfo {volume} {92}},\ \bibinfo
  {pages} {165131} (\bibinfo {year} {2015})}\BibitemShut {NoStop}%
\bibitem [{\citenamefont {Liu}\ \emph {et~al.}(2013)\citenamefont {Liu},
  \citenamefont {Ye},\ and\ \citenamefont {Qi}}]{PhysRevB.87.235306}%
  \BibitemOpen
  \bibfield  {author} {\bibinfo {author} {\bibfnamefont {C.-X.}\ \bibnamefont
  {Liu}}, \bibinfo {author} {\bibfnamefont {P.}~\bibnamefont {Ye}}, \ and\
  \bibinfo {author} {\bibfnamefont {X.-L.}\ \bibnamefont {Qi}},\ }\href
  {\doibase 10.1103/PhysRevB.87.235306} {\bibfield  {journal} {\bibinfo
  {journal} {Phys. Rev. B}\ }\textbf {\bibinfo {volume} {87}},\ \bibinfo
  {pages} {235306} (\bibinfo {year} {2013})}\BibitemShut {NoStop}%
\bibitem [{\citenamefont {Hutasoit}\ \emph {et~al.}(2014)\citenamefont
  {Hutasoit}, \citenamefont {Zang}, \citenamefont {Roiban},\ and\ \citenamefont
  {Liu}}]{PhysRevB.90.134409}%
  \BibitemOpen
  \bibfield  {author} {\bibinfo {author} {\bibfnamefont {J.~A.}\ \bibnamefont
  {Hutasoit}}, \bibinfo {author} {\bibfnamefont {J.}~\bibnamefont {Zang}},
  \bibinfo {author} {\bibfnamefont {R.}~\bibnamefont {Roiban}}, \ and\ \bibinfo
  {author} {\bibfnamefont {C.-X.}\ \bibnamefont {Liu}},\ }\href {\doibase
  10.1103/PhysRevB.90.134409} {\bibfield  {journal} {\bibinfo  {journal} {Phys.
  Rev. B}\ }\textbf {\bibinfo {volume} {90}},\ \bibinfo {pages} {134409}
  (\bibinfo {year} {2014})}\BibitemShut {NoStop}%
\bibitem [{\citenamefont {Grushin}\ \emph {et~al.}(2016)\citenamefont
  {Grushin}, \citenamefont {Venderbos}, \citenamefont {Vishwanath},\ and\
  \citenamefont {Ilan}}]{PhysRevX.6.041046}%
  \BibitemOpen
  \bibfield  {author} {\bibinfo {author} {\bibfnamefont {A.~G.}\ \bibnamefont
  {Grushin}}, \bibinfo {author} {\bibfnamefont {J.~W.~F.}\ \bibnamefont
  {Venderbos}}, \bibinfo {author} {\bibfnamefont {A.}~\bibnamefont
  {Vishwanath}}, \ and\ \bibinfo {author} {\bibfnamefont {R.}~\bibnamefont
  {Ilan}},\ }\href {\doibase 10.1103/PhysRevX.6.041046} {\bibfield  {journal}
  {\bibinfo  {journal} {Phys. Rev. X}\ }\textbf {\bibinfo {volume} {6}},\
  \bibinfo {pages} {041046} (\bibinfo {year} {2016})}\BibitemShut {NoStop}%
\bibitem [{\citenamefont {Weststr\"om}\ and\ \citenamefont
  {Ojanen}(2017)}]{PhysRevX.7.041026}%
  \BibitemOpen
  \bibfield  {author} {\bibinfo {author} {\bibfnamefont {A.}~\bibnamefont
  {Weststr\"om}}\ and\ \bibinfo {author} {\bibfnamefont {T.}~\bibnamefont
  {Ojanen}},\ }\href {\doibase 10.1103/PhysRevX.7.041026} {\bibfield  {journal}
  {\bibinfo  {journal} {Phys. Rev. X}\ }\textbf {\bibinfo {volume} {7}},\
  \bibinfo {pages} {041026} (\bibinfo {year} {2017})}\BibitemShut {NoStop}%
\bibitem [{\citenamefont {Hills}\ \emph {et~al.}(2017)\citenamefont {Hills},
  \citenamefont {Kusmartseva},\ and\ \citenamefont
  {Kusmartsev}}]{PhysRevB.95.214103}%
  \BibitemOpen
  \bibfield  {author} {\bibinfo {author} {\bibfnamefont {R.~D.~Y.}\
  \bibnamefont {Hills}}, \bibinfo {author} {\bibfnamefont {A.}~\bibnamefont
  {Kusmartseva}}, \ and\ \bibinfo {author} {\bibfnamefont {F.~V.}\ \bibnamefont
  {Kusmartsev}},\ }\href {\doibase 10.1103/PhysRevB.95.214103} {\bibfield
  {journal} {\bibinfo  {journal} {Phys. Rev. B}\ }\textbf {\bibinfo {volume}
  {95}},\ \bibinfo {pages} {214103} (\bibinfo {year} {2017})}\BibitemShut
  {NoStop}%
\bibitem [{\citenamefont {Sumiyoshi}\ and\ \citenamefont
  {Fujimoto}(2016)}]{Sumiyoshi_2016}%
  \BibitemOpen
  \bibfield  {author} {\bibinfo {author} {\bibfnamefont {H.}~\bibnamefont
  {Sumiyoshi}}\ and\ \bibinfo {author} {\bibfnamefont {S.}~\bibnamefont
  {Fujimoto}},\ }\href {\doibase 10.1103/PhysRevLett.116.166601} {\bibfield
  {journal} {\bibinfo  {journal} {Phys. Rev. Lett.}\ }\textbf {\bibinfo
  {volume} {116}},\ \bibinfo {pages} {166601} (\bibinfo {year}
  {2016})}\BibitemShut {NoStop}%
\bibitem [{\citenamefont {Huang}\ \emph {et~al.}(2017)\citenamefont {Huang},
  \citenamefont {Zhou},\ and\ \citenamefont {Shen}}]{Huang_2017}%
  \BibitemOpen
  \bibfield  {author} {\bibinfo {author} {\bibfnamefont {Z.-M.}\ \bibnamefont
  {Huang}}, \bibinfo {author} {\bibfnamefont {J.}~\bibnamefont {Zhou}}, \ and\
  \bibinfo {author} {\bibfnamefont {S.-Q.}\ \bibnamefont {Shen}},\ }\href
  {\doibase 10.1103/PhysRevB.96.085201} {\bibfield  {journal} {\bibinfo
  {journal} {Phys. Rev. B}\ }\textbf {\bibinfo {volume} {96}},\ \bibinfo
  {pages} {085201} (\bibinfo {year} {2017})}\BibitemShut {NoStop}%
\bibitem [{\citenamefont {Cortijo}\ \emph
  {et~al.}(2016{\natexlab{a}})\citenamefont {Cortijo}, \citenamefont
  {Ferreir{\'{o}}s}, \citenamefont {Landsteiner},\ and\ \citenamefont
  {Vozmediano}}]{Cortijo_2016}%
  \BibitemOpen
  \bibfield  {author} {\bibinfo {author} {\bibfnamefont {A.}~\bibnamefont
  {Cortijo}}, \bibinfo {author} {\bibfnamefont {Y.}~\bibnamefont
  {Ferreir{\'{o}}s}}, \bibinfo {author} {\bibfnamefont {K.}~\bibnamefont
  {Landsteiner}}, \ and\ \bibinfo {author} {\bibfnamefont {M.~A.~H.}\
  \bibnamefont {Vozmediano}},\ }\href {\doibase 10.1088/2053-1583/3/1/011002}
  {\bibfield  {journal} {\bibinfo  {journal} {2D Materials}\ }\textbf {\bibinfo
  {volume} {3}},\ \bibinfo {pages} {011002} (\bibinfo {year}
  {2016}{\natexlab{a}})}\BibitemShut {NoStop}%
\bibitem [{\citenamefont {Chernodub}\ and\ \citenamefont
  {Vozmediano}(2019)}]{Chernodub_2019}%
  \BibitemOpen
  \bibfield  {author} {\bibinfo {author} {\bibfnamefont {M.~N.}\ \bibnamefont
  {Chernodub}}\ and\ \bibinfo {author} {\bibfnamefont {M.~A.~H.}\ \bibnamefont
  {Vozmediano}},\ }\href {\doibase 10.1103/PhysRevResearch.1.032040} {\bibfield
   {journal} {\bibinfo  {journal} {Phys. Rev. Research}\ }\textbf {\bibinfo
  {volume} {1}},\ \bibinfo {pages} {032040} (\bibinfo {year}
  {2019})}\BibitemShut {NoStop}%
\bibitem [{\citenamefont {Heidari}\ \emph {et~al.}(2019)\citenamefont
  {Heidari}, \citenamefont {Cortijo},\ and\ \citenamefont
  {Asgari}}]{PhysRevB.100.165427}%
  \BibitemOpen
  \bibfield  {author} {\bibinfo {author} {\bibfnamefont {S.}~\bibnamefont
  {Heidari}}, \bibinfo {author} {\bibfnamefont {A.}~\bibnamefont {Cortijo}}, \
  and\ \bibinfo {author} {\bibfnamefont {R.}~\bibnamefont {Asgari}},\ }\href
  {\doibase 10.1103/PhysRevB.100.165427} {\bibfield  {journal} {\bibinfo
  {journal} {Phys. Rev. B}\ }\textbf {\bibinfo {volume} {100}},\ \bibinfo
  {pages} {165427} (\bibinfo {year} {2019})}\BibitemShut {NoStop}%
\bibitem [{\citenamefont {Cortijo}\ \emph
  {et~al.}(2016{\natexlab{b}})\citenamefont {Cortijo}, \citenamefont
  {Kharzeev}, \citenamefont {Landsteiner},\ and\ \citenamefont
  {Vozmediano}}]{Cortijo_PRB}%
  \BibitemOpen
  \bibfield  {author} {\bibinfo {author} {\bibfnamefont {A.}~\bibnamefont
  {Cortijo}}, \bibinfo {author} {\bibfnamefont {D.}~\bibnamefont {Kharzeev}},
  \bibinfo {author} {\bibfnamefont {K.}~\bibnamefont {Landsteiner}}, \ and\
  \bibinfo {author} {\bibfnamefont {M.~A.~H.}\ \bibnamefont {Vozmediano}},\
  }\href {\doibase 10.1103/PhysRevB.94.241405} {\bibfield  {journal} {\bibinfo
  {journal} {Phys. Rev. B}\ }\textbf {\bibinfo {volume} {94}},\ \bibinfo
  {pages} {241405} (\bibinfo {year} {2016}{\natexlab{b}})}\BibitemShut
  {NoStop}%
\bibitem [{\citenamefont {van~der Wurff}\ and\ \citenamefont
  {Cortijo}(2019)}]{van_der_Wurff_2019}%
  \BibitemOpen
  \bibfield  {author} {\bibinfo {author} {\bibfnamefont {E.~C.~I.}\
  \bibnamefont {van~der Wurff}}\ and\ \bibinfo {author} {\bibfnamefont
  {A.}~\bibnamefont {Cortijo}},\ }\href {\doibase
  10.1103/PhysRevResearch.1.033070} {\bibfield  {journal} {\bibinfo  {journal}
  {Phys. Rev. Research}\ }\textbf {\bibinfo {volume} {1}},\ \bibinfo {pages}
  {033070} (\bibinfo {year} {2019})}\BibitemShut {NoStop}%
\bibitem [{\citenamefont {Ghosh}\ \emph {et~al.}(2019)\citenamefont {Ghosh},
  \citenamefont {Sinha}, \citenamefont {Nandy},\ and\ \citenamefont
  {Taraphder}}]{ghosh2019chirality}%
  \BibitemOpen
  \bibfield  {author} {\bibinfo {author} {\bibfnamefont {S.}~\bibnamefont
  {Ghosh}}, \bibinfo {author} {\bibfnamefont {D.}~\bibnamefont {Sinha}},
  \bibinfo {author} {\bibfnamefont {S.}~\bibnamefont {Nandy}}, \ and\ \bibinfo
  {author} {\bibfnamefont {A.}~\bibnamefont {Taraphder}},\ }\href@noop {}
  {\bibfield  {journal} {\bibinfo  {journal} {arXiv preprint arXiv:1911.01130}\
  } (\bibinfo {year} {2019})}\BibitemShut {NoStop}%
\bibitem [{\citenamefont {Jian-Hui}\ \emph {et~al.}(2013)\citenamefont
  {Jian-Hui}, \citenamefont {Hua}, \citenamefont {Qian},\ and\ \citenamefont
  {Jun-Ren}}]{Zhou_2013}%
  \BibitemOpen
  \bibfield  {author} {\bibinfo {author} {\bibfnamefont {Z.}~\bibnamefont
  {Jian-Hui}}, \bibinfo {author} {\bibfnamefont {J.}~\bibnamefont {Hua}},
  \bibinfo {author} {\bibfnamefont {N.}~\bibnamefont {Qian}}, \ and\ \bibinfo
  {author} {\bibfnamefont {S.}~\bibnamefont {Jun-Ren}},\ }\href@noop {}
  {\bibfield  {journal} {\bibinfo  {journal} {Chinese Physics Letters}\
  }\textbf {\bibinfo {volume} {30}},\ \bibinfo {pages} {027101} (\bibinfo
  {year} {2013})}\BibitemShut {NoStop}%
\bibitem [{\citenamefont {Jia}\ \emph {et~al.}(2019)\citenamefont {Jia},
  \citenamefont {Zhang}, \citenamefont {Gao}, \citenamefont {Guo},
  \citenamefont {Yang}, \citenamefont {Hu}, \citenamefont {Bi}, \citenamefont
  {Xiang}, \citenamefont {Liu},\ and\ \citenamefont {Zhang}}]{Jia_2019}%
  \BibitemOpen
  \bibfield  {author} {\bibinfo {author} {\bibfnamefont {H.}~\bibnamefont
  {Jia}}, \bibinfo {author} {\bibfnamefont {R.}~\bibnamefont {Zhang}}, \bibinfo
  {author} {\bibfnamefont {W.}~\bibnamefont {Gao}}, \bibinfo {author}
  {\bibfnamefont {Q.}~\bibnamefont {Guo}}, \bibinfo {author} {\bibfnamefont
  {B.}~\bibnamefont {Yang}}, \bibinfo {author} {\bibfnamefont {J.}~\bibnamefont
  {Hu}}, \bibinfo {author} {\bibfnamefont {Y.}~\bibnamefont {Bi}}, \bibinfo
  {author} {\bibfnamefont {Y.}~\bibnamefont {Xiang}}, \bibinfo {author}
  {\bibfnamefont {C.}~\bibnamefont {Liu}}, \ and\ \bibinfo {author}
  {\bibfnamefont {S.}~\bibnamefont {Zhang}},\ }\href {\doibase
  10.1126/science.aau7707} {\bibfield  {journal} {\bibinfo  {journal}
  {Science}\ }\textbf {\bibinfo {volume} {363}},\ \bibinfo {pages} {148}
  (\bibinfo {year} {2019})}\BibitemShut {NoStop}%
\bibitem [{\citenamefont {Peri}\ \emph {et~al.}(2019)\citenamefont {Peri},
  \citenamefont {Serra-Garcia}, \citenamefont {Ilan},\ and\ \citenamefont
  {Huber}}]{Peri_2019}%
  \BibitemOpen
  \bibfield  {author} {\bibinfo {author} {\bibfnamefont {V.}~\bibnamefont
  {Peri}}, \bibinfo {author} {\bibfnamefont {M.}~\bibnamefont {Serra-Garcia}},
  \bibinfo {author} {\bibfnamefont {R.}~\bibnamefont {Ilan}}, \ and\ \bibinfo
  {author} {\bibfnamefont {S.~D.}\ \bibnamefont {Huber}},\ }\href {\doibase
  10.1038/s41567-019-0415-x} {\bibfield  {journal} {\bibinfo  {journal} {Nature
  Physics}\ }\textbf {\bibinfo {volume} {15}},\ \bibinfo {pages} {357}
  (\bibinfo {year} {2019})}\BibitemShut {NoStop}%
\bibitem [{\citenamefont {Kamboj}\ \emph {et~al.}(2019)\citenamefont {Kamboj},
  \citenamefont {Rana}, \citenamefont {Sirohi}, \citenamefont {Vasdev},
  \citenamefont {Mandal}, \citenamefont {Marik}, \citenamefont {Singh},
  \citenamefont {Das},\ and\ \citenamefont {Sheet}}]{Kamboj_2019}%
  \BibitemOpen
  \bibfield  {author} {\bibinfo {author} {\bibfnamefont {S.}~\bibnamefont
  {Kamboj}}, \bibinfo {author} {\bibfnamefont {P.~S.}\ \bibnamefont {Rana}},
  \bibinfo {author} {\bibfnamefont {A.}~\bibnamefont {Sirohi}}, \bibinfo
  {author} {\bibfnamefont {A.}~\bibnamefont {Vasdev}}, \bibinfo {author}
  {\bibfnamefont {M.}~\bibnamefont {Mandal}}, \bibinfo {author} {\bibfnamefont
  {S.}~\bibnamefont {Marik}}, \bibinfo {author} {\bibfnamefont {R.~P.}\
  \bibnamefont {Singh}}, \bibinfo {author} {\bibfnamefont {T.}~\bibnamefont
  {Das}}, \ and\ \bibinfo {author} {\bibfnamefont {G.}~\bibnamefont {Sheet}},\
  }\href {\doibase 10.1103/PhysRevB.100.115105} {\bibfield  {journal} {\bibinfo
   {journal} {Phys. Rev. B}\ }\textbf {\bibinfo {volume} {100}},\ \bibinfo
  {pages} {115105} (\bibinfo {year} {2019})}\BibitemShut {NoStop}%
\bibitem [{\citenamefont {Zyuzin}(2018)}]{PhysRevB.98.165205}%
  \BibitemOpen
  \bibfield  {author} {\bibinfo {author} {\bibfnamefont {V.~A.}\ \bibnamefont
  {Zyuzin}},\ }\href {\doibase 10.1103/PhysRevB.98.165205} {\bibfield
  {journal} {\bibinfo  {journal} {Phys. Rev. B}\ }\textbf {\bibinfo {volume}
  {98}},\ \bibinfo {pages} {165205} (\bibinfo {year} {2018})}\BibitemShut
  {NoStop}%
\bibitem [{\citenamefont {Zyuzin}\ \emph {et~al.}(2018)\citenamefont {Zyuzin},
  \citenamefont {Silaev},\ and\ \citenamefont {Zyuzin}}]{PhysRevB.98.205149}%
  \BibitemOpen
  \bibfield  {author} {\bibinfo {author} {\bibfnamefont {A.~A.}\ \bibnamefont
  {Zyuzin}}, \bibinfo {author} {\bibfnamefont {M.}~\bibnamefont {Silaev}}, \
  and\ \bibinfo {author} {\bibfnamefont {V.~A.}\ \bibnamefont {Zyuzin}},\
  }\href {\doibase 10.1103/PhysRevB.98.205149} {\bibfield  {journal} {\bibinfo
  {journal} {Phys. Rev. B}\ }\textbf {\bibinfo {volume} {98}},\ \bibinfo
  {pages} {205149} (\bibinfo {year} {2018})}\BibitemShut {NoStop}%
\bibitem [{\citenamefont {Breitkreiz}(2019)}]{breitkreiz2019parabolic}%
  \BibitemOpen
  \bibfield  {author} {\bibinfo {author} {\bibfnamefont {M.}~\bibnamefont
  {Breitkreiz}},\ }\href {https://arxiv.org/abs/1911.08982} {\bibfield
  {journal} {\bibinfo  {journal} {arXiv preprint arXiv:1911.08982}\ } (\bibinfo
  {year} {2019})}\BibitemShut {NoStop}%
\bibitem [{\citenamefont {Duval}\ \emph {et~al.}(2006)\citenamefont {Duval},
  \citenamefont {Horv\'ath}, \citenamefont {Horv\'athy}, \citenamefont
  {Martina},\ and\ \citenamefont {Stichel}}]{Duval_2006}%
  \BibitemOpen
  \bibfield  {author} {\bibinfo {author} {\bibfnamefont {C.}~\bibnamefont
  {Duval}}, \bibinfo {author} {\bibfnamefont {Z.}~\bibnamefont {Horv\'ath}},
  \bibinfo {author} {\bibfnamefont {P.~A.}\ \bibnamefont {Horv\'athy}},
  \bibinfo {author} {\bibfnamefont {L.}~\bibnamefont {Martina}}, \ and\
  \bibinfo {author} {\bibfnamefont {P.~C.}\ \bibnamefont {Stichel}},\ }\href
  {\doibase 10.1103/PhysRevLett.96.099701} {\bibfield  {journal} {\bibinfo
  {journal} {Phys. Rev. Lett.}\ }\textbf {\bibinfo {volume} {96}},\ \bibinfo
  {pages} {099701} (\bibinfo {year} {2006})}\BibitemShut {NoStop}%
\bibitem [{\citenamefont {Roy}\ \emph {et~al.}(2016)\citenamefont {Roy},
  \citenamefont {Juri{\v{c}}i{\'{c}}},\ and\ \citenamefont {Sarma}}]{Roy_2016}%
  \BibitemOpen
  \bibfield  {author} {\bibinfo {author} {\bibfnamefont {B.}~\bibnamefont
  {Roy}}, \bibinfo {author} {\bibfnamefont {V.}~\bibnamefont
  {Juri{\v{c}}i{\'{c}}}}, \ and\ \bibinfo {author} {\bibfnamefont {S.~D.}\
  \bibnamefont {Sarma}},\ }\href {\doibase 10.1038/srep32446} {\bibfield
  {journal} {\bibinfo  {journal} {Scientific Reports}\ }\textbf {\bibinfo
  {volume} {6}} (\bibinfo {year} {2016}),\ 10.1038/srep32446}\BibitemShut
  {NoStop}%
\bibitem [{\citenamefont {Hosur}\ and\ \citenamefont {Qi}(2015)}]{Hosur_2015}%
  \BibitemOpen
  \bibfield  {author} {\bibinfo {author} {\bibfnamefont {P.}~\bibnamefont
  {Hosur}}\ and\ \bibinfo {author} {\bibfnamefont {X.-L.}\ \bibnamefont {Qi}},\
  }\href {\doibase 10.1103/PhysRevB.91.081106} {\bibfield  {journal} {\bibinfo
  {journal} {Phys. Rev. B}\ }\textbf {\bibinfo {volume} {91}},\ \bibinfo
  {pages} {081106} (\bibinfo {year} {2015})}\BibitemShut {NoStop}%
\bibitem [{\citenamefont {Pu}\ \emph {et~al.}(2015)\citenamefont {Pu},
  \citenamefont {Wu},\ and\ \citenamefont {Yang}}]{PhysRevD.91.025011}%
  \BibitemOpen
  \bibfield  {author} {\bibinfo {author} {\bibfnamefont {S.}~\bibnamefont
  {Pu}}, \bibinfo {author} {\bibfnamefont {S.-Y.}\ \bibnamefont {Wu}}, \ and\
  \bibinfo {author} {\bibfnamefont {D.-L.}\ \bibnamefont {Yang}},\ }\href
  {\doibase 10.1103/PhysRevD.91.025011} {\bibfield  {journal} {\bibinfo
  {journal} {Phys. Rev. D}\ }\textbf {\bibinfo {volume} {91}},\ \bibinfo
  {pages} {025011} (\bibinfo {year} {2015})}\BibitemShut {NoStop}%
\bibitem [{\citenamefont {Zyuzin}\ \emph {et~al.}(2012)\citenamefont {Zyuzin},
  \citenamefont {Wu},\ and\ \citenamefont {Burkov}}]{Zyuzin_2012}%
  \BibitemOpen
  \bibfield  {author} {\bibinfo {author} {\bibfnamefont {A.~A.}\ \bibnamefont
  {Zyuzin}}, \bibinfo {author} {\bibfnamefont {S.}~\bibnamefont {Wu}}, \ and\
  \bibinfo {author} {\bibfnamefont {A.~A.}\ \bibnamefont {Burkov}},\ }\href
  {\doibase 10.1103/PhysRevB.85.165110} {\bibfield  {journal} {\bibinfo
  {journal} {Phys. Rev. B}\ }\textbf {\bibinfo {volume} {85}},\ \bibinfo
  {pages} {165110} (\bibinfo {year} {2012})}\BibitemShut {NoStop}%
\bibitem [{\citenamefont {Landsteiner}\ and\ \citenamefont
  {Liu}(2018)}]{Landsteiner_2018}%
  \BibitemOpen
  \bibfield  {author} {\bibinfo {author} {\bibfnamefont {K.}~\bibnamefont
  {Landsteiner}}\ and\ \bibinfo {author} {\bibfnamefont {Y.}~\bibnamefont
  {Liu}},\ }\href {\doibase 10.1016/j.physletb.2018.04.068} {\bibfield
  {journal} {\bibinfo  {journal} {Physics Letters B}\ }\textbf {\bibinfo
  {volume} {783}},\ \bibinfo {pages} {446} (\bibinfo {year}
  {2018})}\BibitemShut {NoStop}%
\end{thebibliography}%

\end{document}